\documentclass[journal]{IEEEtran}

\usepackage{graphicx}
\usepackage{pgfplots}
\usepackage{subfig}
\usepackage{caption}
\usepackage{tikz}
\usepackage{multirow}
\usepackage{soul}
\usepackage{url}
\usepackage{comment}

\usetikzlibrary{shapes,arrows,patterns}
\tikzstyle{startstop} = [rectangle, rounded corners, minimum width=4cm, minimum height=1cm,text centered, text width=4cm, draw=black, fill=red!30]
\tikzstyle{io} = [trapezium, trapezium left angle=70, trapezium right angle=110, minimum width=4cm, minimum height=1cm, text centered, text width=4cm, draw=black, fill=blue!30]
\tikzstyle{process} = [rectangle, minimum width=2cm, minimum height=1cm, text centered, text width=2cm, draw=black, fill=orange!30]
\tikzstyle{decision} = [diamond, minimum width=4.5cm, minimum height=1cm, aspect=4, text centered, draw=black, fill=green!30]
\tikzstyle{arrow} = [thick,->,>=stealth]

\definecolor{appFilesystem}{HTML}{2B83BA}
\definecolor{instanceFilesystem}{HTML}{FDAE61}
\definecolor{instanceCheckpoint}{HTML}{D7191C}
\definecolor{darkGreen}{rgb}{0.0, 0.5, 0.0}

\definecolor{createApp}{HTML}{4D4D4D}
\definecolor{rsyncApp}{HTML}{5DA5DA}
\definecolor{createInstance}{HTML}{FAA43A}
\definecolor{checkpoint}{HTML}{60BD68}
\definecolor{rsyncInstanceFile}{HTML}{F17CB0}
\definecolor{rsyncInstanceCheck}{HTML}{B2912F}
\definecolor{restore}{HTML}{DECF3F}
\definecolor{otherTasks}{HTML}{F15854}

\usepackage{color}

\title{Live Service Migration in Mobile Edge Clouds}
\author{Andrew Machen, Shiqiang Wang, Kin K. Leung, Bong Jun Ko and Theodoros Salonidis \thanks{A. Machen was with Imperial College London and IBM, United Kingdom, when this work was performed. Email: andrew.c.machen@gmail.com}
\thanks{S. Wang, B. J. Ko, and T. Salonidis are with IBM T. J. Watson Research Center, Yorktown Heights, NY, USA, Email: \{wangshiq, bongjun\_ko, \mbox{tsaloni}\}@us.ibm.com}
\thanks{K. K. Leung is with Imperial College London, United Kingdom. Email: kin.leung@imperial.ac.uk }
\thanks{This is the author's version of the paper accepted for publication in IEEE Wireless Communications.  \newline
\textcopyright 2017 IEEE. Personal use of this material is permitted. Permission from IEEE must be obtained for all other uses, in any current or future media, including reprinting/republishing this material for advertising or promotional purposes, creating new collective works, for resale or redistribution to servers or lists, or reuse of any copyrighted component of this work in other works.}
}

\begin{document}

\maketitle
\begin{abstract}
Mobile edge clouds (MECs) bring the benefits of the cloud closer to the user, by installing small cloud infrastructures at the network edge. This enables a new breed of real-time applications, such as instantaneous object recognition and safety assistance in intelligent transportation systems, that require very low latency. One key issue that comes with proximity is how to ensure that users always receive good performance as they move across different locations. Migrating services between MECs is seen as the means to achieve this. This article presents a layered framework for migrating active service applications that are encapsulated either in virtual machines (VMs) or containers. This layering approach allows a substantial reduction in service downtime. The framework is easy to implement using readily available technologies, and one of its key advantages is that it supports containers, which is a promising emerging technology that offers tangible benefits over VMs. The migration performance of various real applications is evaluated by experiments under the presented framework. Insights drawn from the experimentation results are discussed.
\end{abstract}
\begin{IEEEkeywords}
Cloudlet, containers, edge/fog computing, service migration, virtualization
\end{IEEEkeywords}

\section{Introduction}

Cloud-based mobile applications have become increasingly popular over the recent years \cite{dinh2013survey}. One key issue therein is to ensure that services are always delivered with good performance. 
The current centralized structure of the cloud has led to a generally large geographical separation between the users and the cloud infrastructure. In such a setting, end-to-end communication between user and cloud can involve many network hops resulting in high latency; the ingress bandwidth to the cloud may also suffer from saturation as the cloud infrastructure is accessed on a many-to-one basis \cite{m.satyanarayanan2015}.

\begin{figure}
\centering\includegraphics[width=0.9\columnwidth]{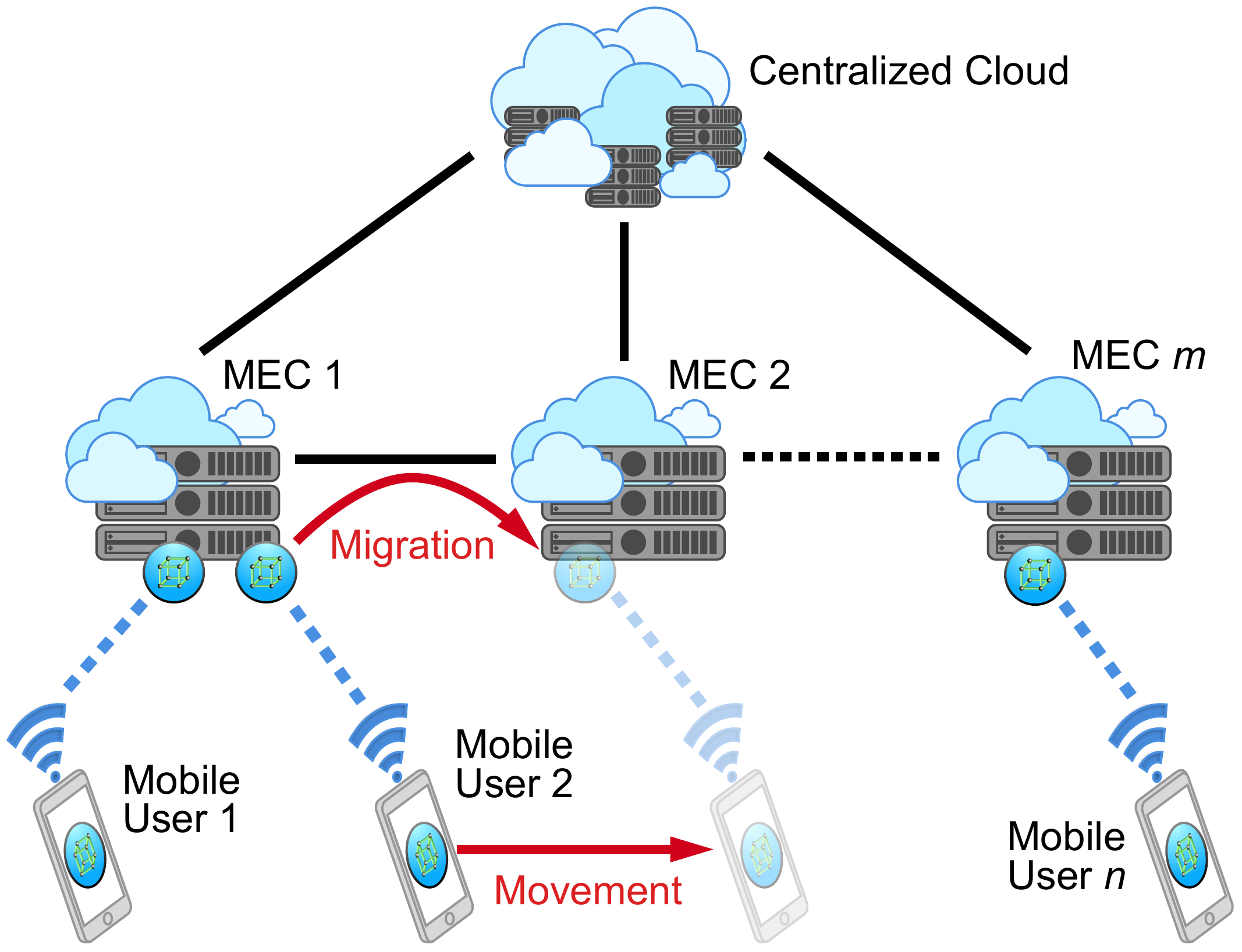}
\caption{Mobile edge cloud (MEC) architecture}
\label{fig:architecture}
\end{figure}

A promising approach for resolving the above problems is to install computing infrastructures at the network edge. 
Particularly for real-time applications such as instantaneous object recognition \cite{ha2013WearableCognitiveAssistance} and safety assistance in intelligent transportation systems (ITS) \cite{VehicularFogComputing}, service applications have to remain in relatively close proximity to their end users in order to ensure low latency and high bandwidth connectivity.
This is captured by the newly emerged concept of \emph{mobile edge clouds (MECs)} \cite{ETSIWhitepaper}, as well as similar concepts such as cloudlet \cite{m.satyanarayanan2015}, fog computing \cite{VehicularFogComputing}, follow-me cloud \cite{FollowMeMagazine}, mobile micro-cloud \cite{wang2015dynamic}, and small cell cloud \cite{becvar2014pimrc}. While these different concepts may propose slightly different implementations, they all suggest placing small cloud infrastructures at the network edge so that users can have seamless connection to cloud services.
In particular, MECs are typically placed only one or a few network hops away from the mobile user, thus the communication latency can be kept very low.
The ingress traffic into the backhaul network can also be reduced, because large amounts of data can be processed directly at the edge.
It is envisioned that MECs will co-exist and work in unison with the existing centralized cloud that we have today, as shown in Figure \ref{fig:architecture}.

To maintain the benefits of running services close to the user, when a mobile user moves away from its original location, its service may need to be \emph{migrated} to a new MEC server that is near the current user location \cite{FollowMeMagazine}. This article focuses on systems aspects of live service migration in MEC environments.
The main challenge is how to maintain relatively low service downtime and overall migration time. We address this challenge with a layered framework which decomposes a cloud application into multiple layers so that only those layers that are missing at the destination are transferred. Our framework applies to both virtual machines and containers and can be readily implemented with existing tools. 
To the best of our knowledge, this is the first systematic study on live migration of MEC applications in a container-based environment.

In the rest of this article, we first summarize the motivation and background, then propose our layered framework for live service migration and its experimentation results.

\section{Live Service Migration: Motivation and Background}

The need for live service migration in MECs can be illustrated with the example shown in Figure~\ref{fig:architecture}. Here, mobile user~2 is initially connected to its service running on MEC~1. The direct connection between user~2 and MEC~1 ensures low-latency access to the service. However, after some time, user~2 moves to a location that does not have direct connection to MEC~1; it has direct connection to MEC~2 instead. As shown in \cite{ha2015adaptive}, connecting to MEC~1 in such cases would incur a significantly higher latency (due to backhaul network communication) than connecting to MEC~2 directly. It can be therefore beneficial to migrate user~2's service from MEC~1 to MEC~2, so that user~2 can continue enjoying low-latency access to its service.
Such live migrations may be frequently required, especially in vehicular applications where it is likely that the users have high mobility. 

Service migration can be classified into stateful and stateless migrations. Stateless migration does not move application running states, it only redirects the user requests to a new server with a separate instance of the service running. This is applicable for applications which do not keep states for users. However, for interactive services that are becoming increasingly popular today, such as active safety warning, mobile multimedia, and mobile online gaming, it is very likely that the application needs to keep some state for each user.  We therefore focus on \emph{stateful} migration in this article, which involves moving running states of applications. We consider the stateful migration of a guest operating system (OS) hosting service applications, where a user receives service for a continuous period of time, and the service application may need to keep some internal state for the user (e.g., some intermediate data processing results). After migration is completed, programs resume exactly where they left off before migration, thus the migration is classified as \emph{live}. The user starts to receive service before migration occurs, and it continues receiving service after migration.

\subsubsection*{Optimization of Migration Decisions}
Migration can incur service interruption as well as computation and communication resource overheads. Therefore, the decision on whether, when, and where to migrate depends on many aspects, such as user mobility, communication channel characteristics, resource availability at MECs, etc., which is a sophisticated optimization problem. 
In essence, there is a trade-off between the cost of migration and the benefit after migration. Algorithms for making migration decisions need to balance this trade-off. They usually need to predict the future service demands with some accuracy or buffer service requests in queues so that they can be served in batches possibly after migration.
Readers are referred to \cite{wang2015dynamic,ceselli2015cloudlet} on how live migration decisions can be formulated as optimization problems and how to solve these problems.

\subsubsection*{Execution of Live Migration }

An active systems research challenge is how to efficiently execute live migration in a practical cloud system containing MECs. 
We first recall that services running on a cloud platform are likely to have OS and application dependencies that need to be met by their host system. 
Therefore, a service application is often encapsulated into its own self-sufficient and pre-configured environment for easy distribution. 
Current examples of such an environment are the well established hypervisor-based \emph{virtual machines (VMs)}, or the relatively new technique, \emph{containers}.
Both technologies allow the creation and running of multiple isolated guest OSs on top of a host OS. The main difference between the two technologies is that VMs fully emulate the OS kernel and hardware, whereas containers directly share the hardware and kernel with their host machines. As a result, containers occupy much less resources and have lower virtualization overhead than VMs, but are less adaptable, e.g., a Linux container cannot run on a Windows server.

Recent effort towards the implementation of service migration in MEC environments has focused on VM migration \cite{ha2015adaptive}. 
Container migration is a relatively new area which has not been systematically studied in the literature. As containers usually have a much smaller size than VMs, it can be very beneficial to run container-based applications on MECs that have limited storage and processing capability. 
Thus, a natural question to ask is how to support the live migration of containers. We particularly would like to support container live migration without drastically changing the existing container implementations, so that minimal effort is required to add this functionality to existing systems.

We should also note that there are existing VM live migration methods for cloud environments \cite{LiveMigrationSurvey}. However, most of them are built for data centers, requiring the use of storage area networks (SANs) and shared storage. Moreover, these methods are usually specific to the underlying virtualization technology. The method presented in this article is designed to work over wide area networks (WANs) which is envisioned to be the way that MECs are interconnected, and it is a generic mechanism that applies for different types of containers and VMs. 

\section{A Layered Framework for Live Migration}
We present a generic layered migration framework using incremental file synchronization, which \emph{works with both containers and VMs}. 
A benefit of this framework is that it is built based on readily available functionalities in most container and VM technologies that are popular today, which means that one does not need modify the internals of container and VM implementation in order to apply this framework.

We focus on LXC (\url{linuxcontainers.org}) and KVM (\url{www.linux-kvm.org}) as representative technologies for containers and virtual machines, respectively. LXC and KVM are chosen for their popularity, and their ability to run Linux-built applications without modification. 
We note that this article focuses on the migration of back-end application components between different MECs, where LXC and KVM are applicable because the server often has a Linux-based operating system. 
The core idea of the method we use for live migration is derived from \cite{tychoandersen2014}, which proposed an LXC live migration mechanism. We have largely extended \cite{tychoandersen2014} so that our approach works with multiple layers (see below), applies (with minor alterations) to the live migration of KVM and undoubtedly other container and VM technologies as well.

In the following, we first describe a basic procedure that we have developed for performing stateful live migrations in MEC environments. Then we describe our layered framework built on top of this basic procedure that optimizes live migration time further.

\subsection{Basic Procedure of Stateful Live Migration in MECs}

To migrate an application, the in-memory state of a running guest OS is recorded, transferred, and then recreated at the destination. The in-memory state includes the applications, system processes, and resources currently loaded into memory for quick access. It contains the progress (state) of running applications, including any data that the application is currently working on. Transferring the in-memory state makes it possible to restore an application exactly from where it was suspended.

The migration framework we present uses accessible tools that already exist in container and VM technology. In LXC, this goes by the name of \emph{checkpointing}, and in KVM, \emph{saving}. Both methods suspend the guest OS (at which point the service is temporarily stopped) and save down the in-memory state of the guest OS into one or more files that can be easily transferred. Complementary tools exist to restore the guest OS from their checkpoint (or save).

After the in-memory state is saved into files, the next step is to use a file transfer protocol to transfer the guest OS's filesystem (i.e., all files saved on the hard disk of the guest OS) and saved in-memory state\footnote{Note that besides the filesystem and in-memory state, we also need to transfer any additional files/data related to the container or VM virtualization itself. We do not specifically discuss those additional portions of data, and only emphasize on the main part of data being transferred. Also, the filesystem and in-memory state may be saved into a single file (in the case of a VM image) or multiple files (in the case of containers). We do not separately discuss them since \textit{rsync} can find differences in two single files as well as multiple files.}. For this we use incremental file synchronization, namely in the form of \textit{rsync} -- a well-known file syncing tool (\url{rsync.samba.org}), to compress and transfer files. A major difference of incremental file synchronization over other basic file transfer protocols is its ability to identify and transfer only those files, or parts thereof, which are different from those already located at the destination. This can substantially reduce the amount of data that needs to be transferred, particularly with our layered framework presented next.

\subsection{Layering}

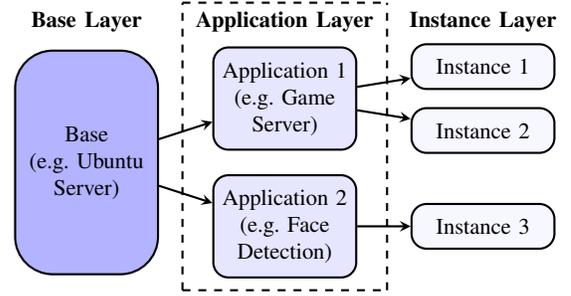
\begin{figure}
\centering
\begin{tikzpicture}[node distance=3.1cm, thick, scale=0.85, every node/.style={scale=0.85}]
\node (base) [process, minimum height=3.5cm, align=center, rounded corners=0.5cm, fill=blue!30] {Base\\ (e.g. Ubuntu Server)};
\node (app1) [process, right of=base, yshift=1cm, minimum height=1.6cm, align=center, rounded corners=0.2cm, fill=blue!10] {Application 1\\ (e.g. Game Server)};
\node (app2) [process, right of=base, yshift=-1cm, align=center, minimum height=1.6cm, rounded corners=0.2cm, fill=blue!10] {Application 2\\ (e.g. Face Detection)};
\node (instance1) [process, right of=app1, yshift=0.5cm, minimum height=0.7cm, rounded corners=0.2cm, fill=blue!3] {Instance 1};
\node (instance2) [process, right of=app1, yshift=-0.5cm, minimum height=0.7cm, rounded corners=0.2cm, fill=blue!3] {Instance 2};
\node (instance3) [process, right of=app2, minimum height=0.7cm, rounded corners=0.2cm, fill=blue!3] {Instance 3};
\node (baseLayer) [draw=none, above of=base, yshift=-0.9cm, font=\bf] {Base Layer};
\node (appLayer) [draw=none, above of=app1, yshift=-1.9cm, font=\bf] {Application Layer};
\node (instanceLayer) [draw=none, above of=instance1, yshift=-2.4cm, font=\bf] {Instance Layer};
\draw [dashed] (1.5,-2) -- (1.5,2.5);
\draw [dashed] (1.5,2.5) -- (4.7,2.5);
\draw [dashed] (4.7,-2) -- (4.7,2.5);
\draw [dashed] (1.5,-2) -- (4.7,-2);
\draw [arrow] (base) -- (app1);
\draw [arrow] (base) -- (app2);
\draw [arrow] (app1) -- (instance1);
\draw [arrow] (app1) -- (instance2);
\draw [arrow] (app2) -- (instance3);
\end{tikzpicture}
\caption{The three-layer model, where the dashed box outlines the additional building block (i.e., the application layer) that does not exist in the two layer model.}
\label{fig:threeLayerModel}
\end{figure}

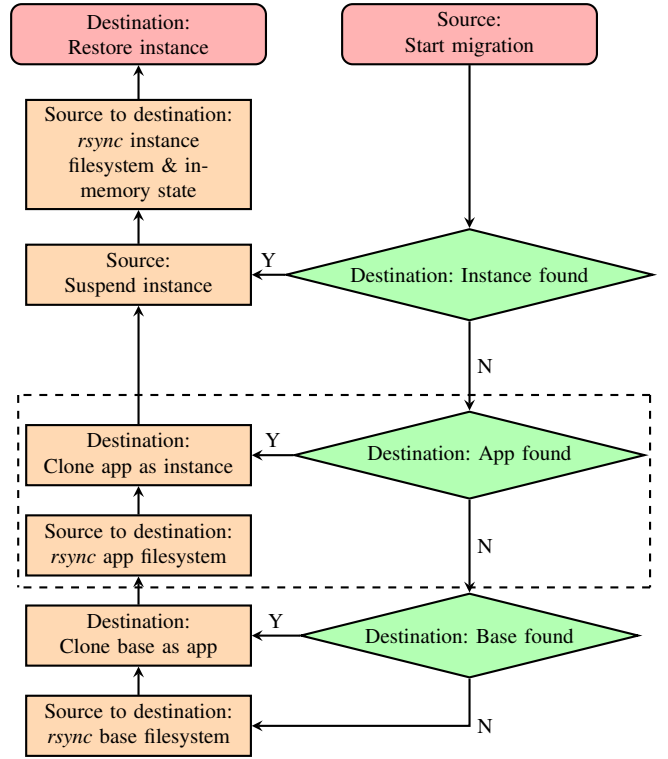
\begin{figure}
\centering
\begin{tikzpicture}[node distance=2cm, thick, scale=0.8, every node/.style={scale=0.8}]
\node (start) [startstop] {Source:\\ Start migration};
\node (instanceFound) [decision, below of=start, yshift=-2cm] {Destination: Instance found};
\node (appFound) [decision, below of=instanceFound, yshift=-1cm] {Destination: App found};
\node (baseFound) [decision, below of=appFound, yshift=-1cm] {Destination: Base found};
\node (cloneBase) [process, minimum width=3.5cm, text width=3.5cm,left of=baseFound, xshift=-3.5cm] {Destination:\\ Clone base as app};
\node (rsyncBase) [process, minimum width=3.5cm, text width=3.5cm,below of=cloneBase, yshift=0.5cm] {Source to destination:\\ \textit{rsync} base filesystem};
\node (cloneApp) [process, minimum width=3.5cm, text width=3.5cm,left of=appFound, xshift=-3.5cm] {Destination:\\ Clone app as instance};
\node (rsyncApp) [process, minimum width=3.5cm, text width=3.5cm,above of=cloneBase, yshift=-0.5cm] {Source to destination:\\ \textit{rsync} app filesystem}; 
\node (checkpoint) [process, minimum width=3.5cm, text width=3.5cm,left of=instanceFound, xshift=-3.5cm] {Source:\\ Suspend instance};
\node (rsyncInstance) [process, minimum width=3.5cm, text width=3.5cm, above of=checkpoint, yshift=-0cm, align=center] {Source to destination:\\ \textit{rsync} instance filesystem \& in-memory state};
\node (end) [startstop, left of=start, xshift=-3.5cm] {Destination:\\ Restore instance};
\draw [arrow] (start) -- (instanceFound);
\draw [arrow] (instanceFound) -- node[anchor=west] {N} (appFound);
\draw [arrow] (appFound) -- node[anchor=west] {N} (baseFound);
\draw [arrow] (baseFound) |- node[anchor=west] {N} (rsyncBase);
\draw [arrow] (instanceFound) -- node[anchor=south] {Y} (checkpoint);
\draw [arrow] (appFound) -- node[anchor=south] {Y} (cloneApp);
\draw [arrow] (baseFound) -- node[anchor=south] {Y} (cloneBase);
\draw [arrow] (rsyncBase) -- (cloneBase);
\draw [arrow] (cloneBase) -- (rsyncApp);
\draw [arrow] (rsyncApp) -- (cloneApp);
\draw [arrow] (cloneApp) -- (checkpoint);
\draw [arrow] (checkpoint) -- (rsyncInstance);
\draw [arrow] (rsyncInstance) -- (end);
\draw [dashed] (3,-6) -- (3,-9.2);
\draw [dashed] (-7.5,-6) -- (-7.5,-9.2);
\draw [dashed] (3,-9.2) -- (-7.5,-9.2);
\draw [dashed] (3,-6) -- (-7.5,-6);
\end{tikzpicture}
\caption{Flow chart of the migration mechanism for live-migrating with the three-layer model, where \textit{rsync} is used for incremental file synchronization. The dashed box outlines the additional building block (i.e., operations related to the application layer) that does not exist in the two layer model.}
\label{fig:migrationMechanism}
\end{figure}

The problem with migrating an encapsulated service directly using the above basic live migration procedure is that the ``package'' contains the guest OS, virtualization data, and, for VMs, the system kernel, which are required to make the service self-contained. A base package with no services installed can have a size that tips the scale at about $400$~MB and $2.7$~GB for LXC and KVM, respectively. Our experiments using the live migration method discussed above have shown that, for a $100$~Mbps bandwidth connection, the average migration time for a base package is $25$~seconds and $160$~seconds for LXC and KVM, respectively, during which time the service is down and may appear as an unresponsive or frozen application.

In our approach, we aim to reduce service downtime (i.e., time of service interruption) and overall migration time (i.e., time from the beginning till the end of the whole migration process) through the use of layers. Abstractly, we can separate the base package (that includes the guest OS, kernel, etc., but with no applications installed) into a layer separate and distinct from its service applications. We call this package the \emph{base layer}. This base layer is generic and all MECs shall have a copy of it. The service applications and their running states are placed within a separate layer called the \emph{instance layer}. Assuming the base layer exists at the destination, when we want to migrate a service, we first suspend the service and then transfer only the instance layer to the destination. A running service can be reconstructed from a combination of the base and instance layers. By removing the need to transfer the base package in every migration, we can drastically cut down the amount of data that needs to be transferred, and in turn reduce the service downtime and overall migration time.

We extend this \emph{two-layer} approach further, by splitting out the application from the instance layer into an intermediate \emph{application layer} that contains an idle version of the application and any application-specific data. The instance layer now only needs to contain the running state (i.e., in-memory state) of a service. When we want to migrate a service, the application layer can be migrated first whilst the service is running, then the service is suspended and the instance layer is transferred to the destination. A running service can be reconstructed from a combination of the base layer, instance layer, and associated application layer. By doing this, we are able to transfer the majority of an encapsulated service's program and data before suspending the service, leaving only the running state transfer to count towards the service downtime. This three-layer setup is shown in Figure~\ref{fig:threeLayerModel}, and is referred to as the \emph{three-layer model}.

The intermediate application layer of the three-layer model has advantages other than improvements to service downtime. A distinct application layer allows the application and its related data to be distributed independently of a running service. Through smart pre-distribution of services, possibly by application caching based on demand prediction, the overall migration time can be reduced to only the time it takes to migrate the instance layer. 

In the current implementation, a layer contains its unique data (including files and possibly in-memory state) plus all the data from its preceding layers. We achieve pseudo-incremental layering for migration, by first cloning a lower layer (e.g., application layer), and then using incremental file synchronization to transfer only the difference between that and the higher layer (e.g., instance layer) we want to recreate. This process of cloning and incremental file synchronization is repeated depending on what layers are missing at the destination, the complete migration process is shown in Figure~\ref{fig:migrationMechanism}.
The time to start the migration and the destination of migration can be determined by an optimization algorithm, see \cite{wang2015dynamic,ceselli2015cloudlet} for details.

\section{Experimentation Results and Discussions}

We study the live migration performance of our layered framework for a variety of applications.
We ran experiments in three ``host'' VMs connected to an emulation framework, where the emulation framework is the open-source Common Open Research Emulator (CORE) \cite{core}. The host VMs were each given $2$ virtual CPU cores and $2$~GB of virtual memory from a physical machine with $2.6$~GHz Intel Core i7 and a total of $16$~GB $1600$~MHz DDR3 memory. Two host VMs acted as MECs, between which migration was carried out, and the third host VM acted as a user requiring MEC service. 
Unless otherwise specified, the bandwidth between MECs was configured as $100$~Mbps with only system-inherited latency and jitter. The connection between each MEC and the user was configured with $100$~Mbps bandwidth, $25$~ms latency, and $5$~ms jitter. These connection specifications are typical for wide-area wired networks and local wireless networks, on which MECs and their users operate. They are much more inferior than network connections in data centers.
Migration times were calculated by a migration script at each stage of the migration process. 
We use \textit{rsync} for incremental file synchronization and the amount of transferred data was measured using Wireshark.

Nested KVM and LXC, which contain running applications, were run inside the two host VMs that mimic MECs. The guest OS of the nested KVM and LXC was Ubuntu 15.10. Note that the base OS image sizes for KVM and LXC are different due to their different virtualization mechanisms, but this has no bearing in our results since, in our experiments, the base layer resides in every MEC and is not included in the migration.
As we see below, the installation footprint of applications can be larger for LXC than for KVM, because the Ubuntu installation of LXC has fewer packages included as standard. As such, the installation of any missing packages is counted into the LXC application size.

We studied the migration of the following applications using our migration framework presented above.
\begin{enumerate}
\item \emph{Game Server} runs the \textit{sauerbraten-server} package, a server for the online game Cube 2. It sends and receives regular packets related to player location and other match statistics. The installation footprint is very small ($0.7$~MB), as is the memory requirement (approximately $1$~MB).

\item \emph{RAM Simulation} is a simple script (approximately $0.1$~MB) that consumes a large amount of RAM, where the exact RAM consumption is user-defined and the RAM contents keep changing over time. It represents memory intensive applications, such as those that process large data sets or perform complex calculations (e.g., big data analysis, training of deep neural networks, etc.). Unless otherwise specified, RAM utilization is maintained at around $330$~MB. 

\item \emph{Video Streaming} uses the \textit{vlc-nox} package (approximately $280$~MB for LXC, $230$~MB for KVM) to stream video to a user. A $50$~MB video file is stored with the application at the MEC. Video Streaming has a low memory requirement (approximately $30$~MB). 

\item \emph{Face Detection} uses the \textit{OpenCV} library to process an incoming video stream. It detects faces in the video received, and sends the detection result, embedded into each video frame, back to the user. This application has a very large installation footprint (approximately $655$~MB for LXC, $565$~MB for KVM), and a moderate memory requirement of approximately $100$~MB.

\item \emph{No Application} is a guest OS with no applications installed, and therefore requires no additional resources. It represents the overhead, or minimum bound, on migration.
\end{enumerate}

\begin{table*}
\caption{Migration results for two-layer and three-layer configurations.}
\label{tab:threeLayers}
\centering
\small
\subfloat{
\begin{tabular}{|p{4.8cm}|p{1.8cm}|l|l|l|p{1.7cm}|p{1.7cm}|}
\cline{1-4} \cline{6-7}
LXC Total Migration Time & \multicolumn{1}{c|}{\multirow{2}{*}{2 layers}} & \multicolumn{1}{c|}{\multirow{2}{*}{\begin{tabular}[c]{@{}c@{}}3 layers\\ (app not found)\end{tabular}}} & \multicolumn{1}{c|}{\multirow{2}{*}{\begin{tabular}[c]{@{}c@{}}3 layers\\ (app found)\end{tabular}}} &  & \multicolumn{2}{l|}{LXC Service Downtime} \\ \cline{1-1} \cline{6-7} 
LXC Total Data Transferred & \multicolumn{1}{c|}{} & \multicolumn{1}{c|}{} & \multicolumn{1}{c|}{} &  & \multicolumn{1}{c|}{2 layers} & \multicolumn{1}{c|}{3 layers} \\ \cline{1-4} \cline{6-7} 
\multirow{2}{*}{No Application} & 6.5 s & 11.0 s & 6.3 s &  & 2.0 s & 2.0 s \\
 & 1.4 MB & 1.9 MB & 1.4 MB &  &  &  \\ \cline{1-4} \cline{6-7} 
\multirow{2}{*}{Game Server} & 7.3 s & 10.9 s & 6.4 s &  & 3.0 s & 2.0 s \\
 & 2.2 MB & 2.7 MB & 1.6 MB &  &  &  \\ \cline{1-4} \cline{6-7} 
\multirow{2}{*}{RAM Simulation} & 20.2 s & 27.2 s & 19.8 s &  & 15.4 s & 15.3 s \\
 & 97.1 MB & 97.6 MB & 97.1 MB &  &  &  \\ \cline{1-4} \cline{6-7} 
\multirow{2}{*}{Video Streaming} & 27.5 s & 37.3 s & 8.5 s &  & 23.2 s & 3.3 s \\
 & 180.2 MB & 184.6 MB & 7.4 MB &  &  &  \\ \cline{1-4} \cline{6-7} 
\multirow{2}{*}{Face Detection} & 52.0 s & 70.1 s & 15.5 s &  & 47.6 s & 3.7 s \\
 & 363.1 MB & 365.0 MB & 10.0 MB &  &  &  \\ \cline{1-4} \cline{6-7} 
\end{tabular}
}
\hfill
\subfloat{
\begin{tabular}{|p{4.8cm}|p{1.8cm}|l|l|l|p{1.7cm}|p{1.7cm}|}
\cline{1-4} \cline{6-7}
KVM Total Migration Time & \multicolumn{1}{c|}{\multirow{2}{*}{2 layers}} & \multicolumn{1}{c|}{\multirow{2}{*}{\begin{tabular}[c]{@{}c@{}}3 layers\\ (app not found)\end{tabular}}} & \multicolumn{1}{c|}{\multirow{2}{*}{\begin{tabular}[c]{@{}c@{}}3 layers\\ (app found)\end{tabular}}} &  & \multicolumn{2}{l|}{KVM Service Downtime} \\ \cline{1-1} \cline{6-7} 
KVM Total Data Transferred & \multicolumn{1}{c|}{} & \multicolumn{1}{c|}{} & \multicolumn{1}{c|}{} &  & \multicolumn{1}{c|}{2 layers} & \multicolumn{1}{c|}{3 layers} \\ \cline{1-4} \cline{6-7} 
\multirow{2}{*}{No Application} & 81.5 s & 141.8 s & 79.7 s &  & 55.8 s & 56.1 s \\
 & 65.3 MB & 65.9 MB & 65.2 MB &  &  &  \\ \cline{1-4} \cline{6-7} 
\multirow{2}{*}{Game Server} & 84.7 s & 142.0 s & 80.9 s &  & 60.2 s & 58.7 s \\
 & 71.5 MB & 72.3 MB & 69.7 MB &  &  &  \\ \cline{1-4} \cline{6-7} 
\multirow{2}{*}{RAM Simulation} & 95.9 s & 152.2 s & 93.0 s &  & 72.5 s & 72.0 s \\
 & 170.0 MB & 178.4 MB & 170.4 MB &  &  &  \\ \cline{1-4} \cline{6-7} 
\multirow{2}{*}{Video Streaming} & 120.9 s & 189.7 s & 86.4 s &  & 93.9 s & 61.3 s \\
 & 251.3 MB & 270.5 MB & 87.3 MB &  &  &  \\ \cline{1-4} \cline{6-7} 
\multirow{2}{*}{Face Detection} & 381.2 s & 558.3 s & 152.3 s &  & 355.0 s & 107.5 s \\
 & 1,027.3 MB & 1,033.7 MB & 99.0 MB &  &  &  \\ \cline{1-4} \cline{6-7} 
\end{tabular}
}
\end{table*}

\begin{figure*}
	\centering
	\begin{tikzpicture}
		\begin{axis}[
			axis y line=middle,
    		xbar stacked,
    		ytick=data,
		    axis x line*=bottom,
		    tick label style={font=\footnotesize},
		   	legend columns=4,
		    legend style={font=\footnotesize, at={(0.5,1.3)},anchor=north},
		    label style={font=\footnotesize},
		    width=0.9\textwidth,
		    bar width=6mm,
	        xlabel={LXC Migration Time [s]},
		    yticklabels={No Application, Game Server, RAM Intensive App, 
		    			 Video Streaming, Face Detection},
		    y=8mm,
		    enlarge y limits={abs=0.625},
		    axis background/.style={fill=gray!10},
			]
			
			\addplot[createApp,fill=createApp] coordinates
			{(2.87,0) (2.87,1) (5.06,2) (4.55,3) (5.12,4)};
			\addplot[rsyncApp,fill=rsyncApp] coordinates
			{(0.60,0) (1.07,1) (0.62,2) (21.57,3) (47.73,4)};
			\addplot[createInstance,fill=createInstance] coordinates
			{(2.93,0) (2.72,1) (2.71,2) (5.67,3) (11.02,4)};
			\addplot[checkpoint,fill=checkpoint,postaction={pattern=north east lines}] coordinates
			{(0.28,0) (0.28,1) (0.78,2) (0.32,3) (0.38,4)};
			\addplot[rsyncInstanceFile,fill=rsyncInstanceFile,postaction={pattern=north east lines}] coordinates
			{(0.93,0) (0.86,1) (1.31,2) (1.06,3) (1.36,4)};
			\addplot[rsyncInstanceCheck,fill=rsyncInstanceCheck,postaction={pattern=north east lines}] coordinates
			{(0.53,0) (0.52,1) (13.04,2) (1.53,3) (1.61,4)};
			\addplot[restore,fill=restore,postaction={pattern=north east lines}] coordinates
			{(0.39,0) (0.47,1) (1.21,2) (0.50,3) (0.50,4)};
			\addplot[otherTasks,fill=otherTasks] coordinates
			{(2.48,0) (2.16,1) (2.43,2) (2.07,3) (2.43,4)};
			
			\legend{Clone base as app, \textit{rsync} app filesystem, Clone app as instance, Suspend instance, \textit{rsync} instance filesystem\,\,\,, \textit{rsync} instance in-memory state, Restore instance, Other remaining tasks}
		\end{axis}
	\end{tikzpicture}
	\begin{tikzpicture}
		\begin{axis}[
			axis y line=middle,
    		xbar stacked,
    		ytick=data,
		    axis x line*=bottom,
		    tick label style={font=\footnotesize},
		    label style={font=\footnotesize},
		    width=0.9\textwidth,
		    bar width=6mm,
		    xlabel={KVM Migration Time [s]},
		    yticklabels={No Application, Game Server, RAM Intensive App, 
		    			 Video Streaming, Face Detection},
		    y=8mm,
		    enlarge y limits={abs=0.625},
    	    axis background/.style={fill=gray!10},
			]
			
			\addplot[createApp,fill=createApp] coordinates
			{(19.54,0) (19.76,1) (18.81,2) (22.46,3) (22.26,4)};
			\addplot[rsyncApp,fill=rsyncApp] coordinates
			{(48.52,0) (45.71,1) (43.37,2) (79.58,3) (380.37,4)};
			\addplot[createInstance,fill=createInstance] coordinates
			{(13.59,0) (14.00,1) (13.44,2) (21.12,3) (43.88,4)};
			\addplot[checkpoint,fill=checkpoint,postaction={pattern=north east lines}] coordinates
			{(1.72,0) (1.93,1) (3.27,2) (2.20,3) (3.19,4)};
			\addplot[rsyncInstanceFile,fill=rsyncInstanceFile,postaction={pattern=north east lines}] coordinates
			{(41.53,0) (42.36,1) (40.98,2) (43.51,3) (86.77,4)};
			\addplot[rsyncInstanceCheck,fill=rsyncInstanceCheck,postaction={pattern=north east lines}] coordinates
			{(10.74,0) (11.77,1) (25.84,2) (15.40,3) (16.07,4)};
			\addplot[restore,fill=restore,postaction={pattern=north east lines}] coordinates
			{(2.37,0) (2.63,1) (2.82,2) (2.18,3) (2.17,4)};
			\addplot[otherTasks,fill=otherTasks] coordinates
			{(3.83,0) (3.83,1) (3.74,2) (3.28,3) (3.60,4)};
		\end{axis}
	\end{tikzpicture}
	\caption{Application migration time broken down by stages, for the three-layer model when the application has not been found at the destination. The hatched stages represent those that contribute towards service downtime.}
	\label{fig:lxcMigrationTimesforApps}
\end{figure*}
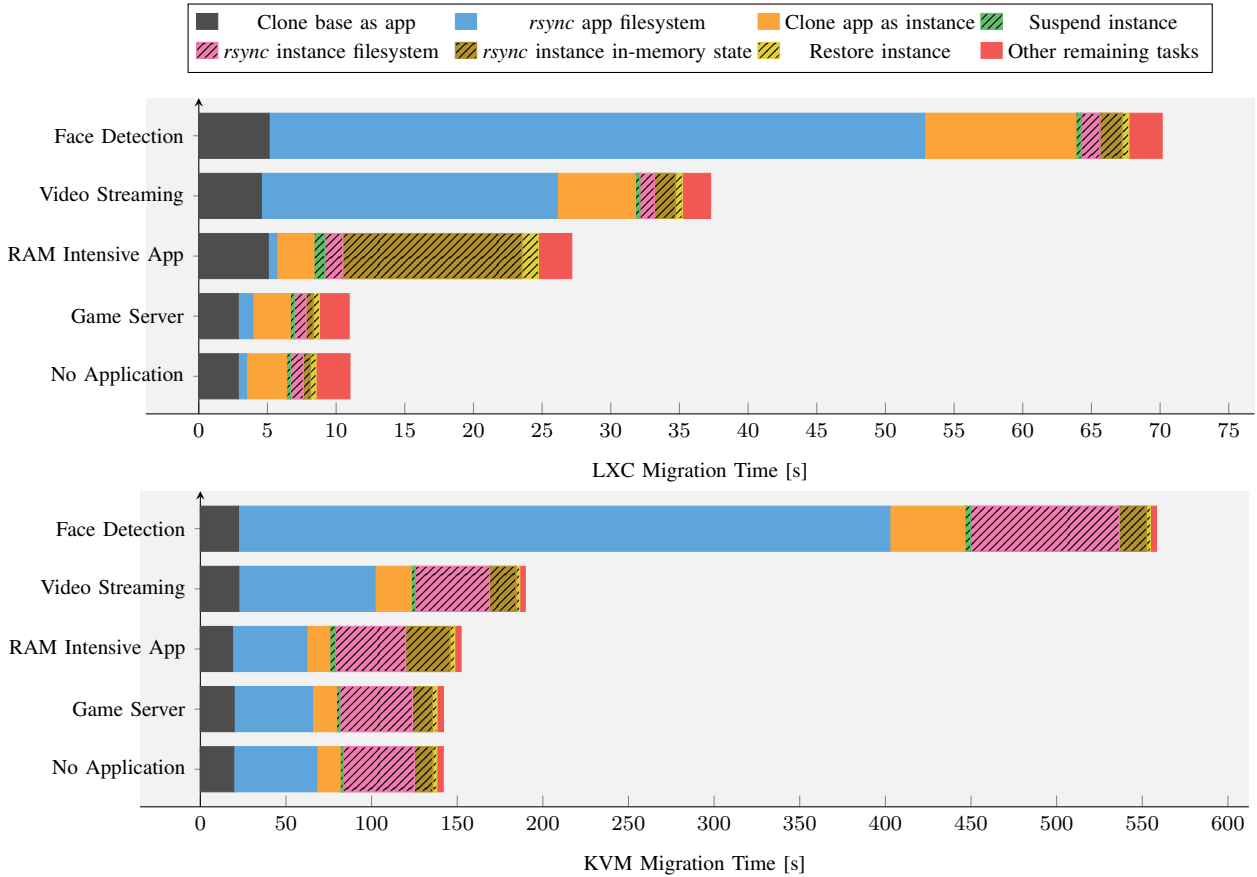

\begin{figure}
	\centering
	\subfloat{
	\begin{tikzpicture}
 		\begin{axis}[
  			width=0.9\columnwidth,
			xlabel={RAM Usage [MBytes]},
			ylabel={Migration Time [s]},
			log ticks with fixed point,
			xmin=0, xmax=600,
			ymin=0, ymax=160,
			axis y line*=none,
		    axis x line*=bottom,
    	    axis background/.style={fill=gray!10},
            tick label style={font=\normalsize},
            legend style={font=\normalsize},
		    label style={font=\normalsize},
        	legend pos= north west,
			]
			\addplot[
				smooth,
				color=appFilesystem,
				mark=*,
				]
				coordinates{
					(20,9.31)(100,12.95)(200,17.91)
					(300,23.40)(400,27.88)(500,32.99)
					(600,39.57)
				};
			\addplot[
				smooth,
				color=instanceCheckpoint,
				mark=*,
				]
				coordinates{
					(380,137.91)(400,139.94)(450,142.93)
					(500,145.02)(550,147.85)(600,151.55)
				};
			\legend{LXC,KVM}
  		\end{axis}
 	\end{tikzpicture}
 	} \hspace{-0.2in}
 	\subfloat{
	\begin{tikzpicture}
 		\begin{axis}[
  			width=0.9\columnwidth,
			xlabel={Bandwidth [Mbps]},
			ylabel={Migration Time [s]},
			xmode=log,
			log ticks with fixed point,
			legend pos=north east,
			xmin=0, xmax=1000,
			ymin=0, ymax=650,
			axis y line*=none,
		    axis x line*=bottom,
    	    axis background/.style={fill=gray!10},
    	    tick label style={font=\normalsize},
            legend style={font=\normalsize},
		    label style={font=\normalsize},
			]
			\addplot[
				smooth,
				color=appFilesystem,
				mark=*,
				]
				coordinates{
					(1,288.14)(2,147.34)(5,63.57)
					(10,35.87)(20,22.87)(50,14.37)
					(100,13.68)(1000,13.64)
				};
			\addplot[
				smooth,
				color=instanceCheckpoint,
				mark=*,
				]
				coordinates{
					(1,624.37)(2,364.04)(5,203.88)
					(10,152.09)(20,125.63)(50,112.26)
					(100,111.51)(1000,111.95)
				};
			\legend{LXC,KVM}
  		\end{axis}
 	\end{tikzpicture} 	
 	}
 	\caption{Total migration time under different RAM usage and bandwidth, for RAM Simulation application.}
 	\label{fig:ramUsage}
\end{figure}
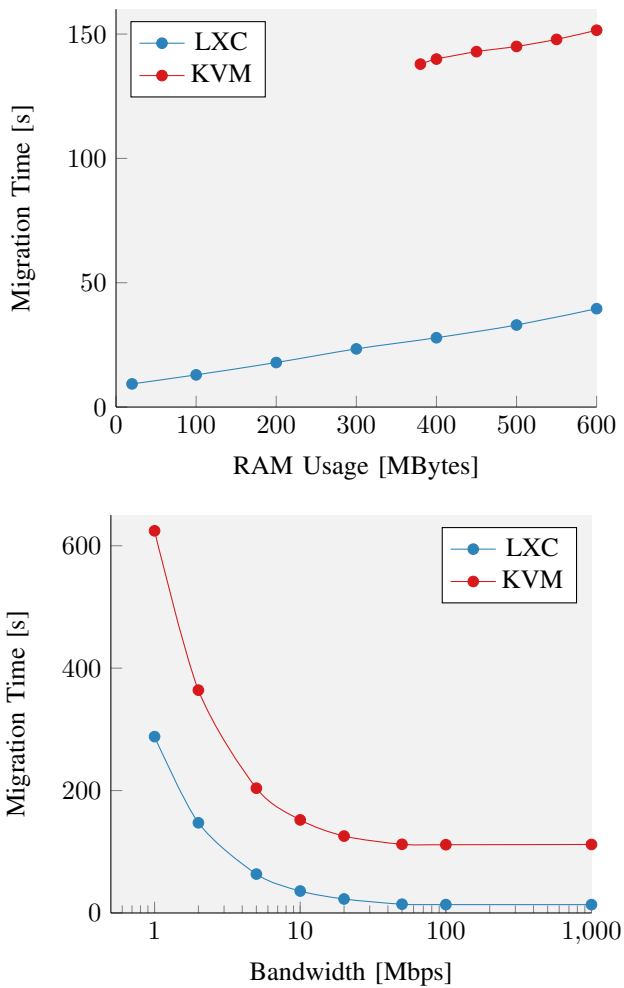

The experimentation results (averaged over $10$ independent experiments in each case) are shown in Table~\ref{tab:threeLayers} and Figures~\ref{fig:lxcMigrationTimesforApps} and \ref{fig:ramUsage}. 
For a particular setup, the service downtime is always smaller than the total migration time, because the total migration time also includes the time taken for cloning and data transfer (during which the service remains running). 
The time required for different migration stages are unequal for different applications, as shown in Figure~\ref{fig:lxcMigrationTimesforApps}, due to the different installation and RAM occupation sizes. This aligns with the layered abstraction discussed earlier. For example, the time required for \textit{rsync}ing the application filesystem is related to the application installation size, and the time required for \textit{rsync}ing the instance in-memory state is related to the application's RAM usage.

\subsection{Container vs. Virtual Machine}

Although it is understood that containers are more lightweight than VMs, a quantitative view on the difference in their migration performance does not exist in the literature. The container-supported migration framework presented above allows us to provide such a quantitative view and draw further insights.

We can see from the experimentation results that for all the example applications shown, LXC has a clear advantage over KVM in terms of total migration time, service downtime, and amount of transferred data.
This is mainly because containers are more compact and the filesystem and in-memory contents of a container is mostly relevant to the application; whereas the filesystem and in-memory contents of a VM can be related to many other background processes irrelevant to the considered application, and \textit{rsync} (or any other incremental file synchronization mechanism) needs to remotely compare a larger amount of data and may not be able to filter out everything that does not belong to the application.

From Figure~\ref{fig:ramUsage} (top), we see that for the RAM Simulation application, the migration times of both LXC and KVM  are approximately linear in RAM utilization. This linear relationship is because the amount of in-memory state data that needs to be transferred is proportional to the RAM usage.
The lower bound on RAM usage represents the overhead required to migrate LXC or KVM. Depending on the application's RAM utilization, we can migrate approximately up to $19\mathsf{x}$ the number of containers compared to VMs within the same amount of time.
That said, the relative advantage of LXC over KVM is reduced as the RAM utilization increases. At $400$~MB of RAM usage, LXC is $5\mathsf{x}$ faster to migrate than KVM; whereas at $600$~MB of RAM usage, LXC is only $4\mathsf{x}$ faster to migrate than KVM.

Figure~\ref{fig:ramUsage} (bottom) shows that the relative advantage of LXC over KVM increases as bandwidth increases. At $1$~Mbps, LXC is $2\mathsf{x}$ faster to migrate than KVM, whereas at $10$~Mbps LXC is $4\mathsf{x}$ faster. We found in our experiments that the data transmission rate is capped by how fast \textit{rsync} is able to compare files and compress data. When the bandwidth is over $50$~Mbps, the migration time remains about the same, where LXC is $8\mathsf{x}$ faster to migrate than KVM.

Containers also have their shortcomings though. By relying on the host system for both the hardware and kernel, they are less adaptable than VMs. They can be nested within VMs to regain their adaptability, but this is likely to degrade performance. Instead, it is advisable that MEC infrastructure is chosen to ensure container compatibility, in as much as the OS (most likely a Linux derivative) being consistent across the MEC network.

There are also considerations other than performance that will need to be addressed, notably security, in order for containers to be recommended over VMs in all scenarios. Security, a pre-requisite for any enterprise software, is a serious concern with containers, as unlike their VM counterparts they are not fully isolated from their host system, and can therefore be more susceptible to attack from a compromised host.

\subsection{Two Layers vs. Three Layers}

From the experimentation results, we see that most applications respond positively towards the three-layer model. For LXC, when the application has been found at the migration target, Table~\ref{tab:threeLayers} and Figure~\ref{fig:lxcMigrationTimesforApps} show that the Video Streaming and Face Detection applications have significant reductions in the total migration time compared to using only two layers, in the orders of $3\mathsf{x}$ for both. Even more significant is the reduction in service downtime for these applications, in the orders of $7\mathsf{x}$ and $13\mathsf{x}$, respectively. This service downtime reduction is particularly important for a seamless experience for the end user. The most significant is the reductions in the amount of data transferred over the network for migration, which are respectively in the orders of $24\mathsf{x}$ and $36\mathsf{x}$. This is particularly important for bandwidth or time constrained connections. Total migration times and data transfers for the Game Server, RAM Simulation, and No Application do not improve significantly, but are also not worse than with two layers when the application is found at the migration target.
The relative reductions in total migration time and data transfers are smaller for KVM, because the application and associated data count for a smaller proportion of the total size of a KVM virtual machine.

We note that Video Streaming and Face Detection both have large filesystems compared to RAM Simulation and Game Server, and it is from this the three-layer model derives its benefit. The larger the application or associated data, the greater the benefit from introducing an intermediate layer. This can be seen in Figure~\ref{fig:lxcMigrationTimesforApps}, which shows how the migration of application data can account for a significant portion of the total time it takes to migrate an application.

The three-layer model does have a downside, in that if the application has not been found at the migration destination, the total migration time can be longer than with only two layers, because additional time is required for cloning the application into an instance and \textit{rsync}ing the third layer (see Figure~\ref{fig:lxcMigrationTimesforApps}). However, the service downtime (which is usually the more critical factor for MEC applications) remains the same no matter whether the application is found at the destination or not, because the service is suspended only after the application layer (containing an idle application) has been transferred. 

Therefore, it is apparent that the benefits of the three-layer model need to be balanced with any trade-offs incurred. Factors such as how frequently the application is used and how much service downtime impacts the user experience should be taken into account when deciding whether to use two or three layers. 
Such decisions have to be made based on application characteristics. 
For example, it may be better for applications with small installation sizes and intensive RAM utilization to use only two layers for quicker migration.

A practical implementation of an MEC system can support both two-layer and three-layer models, since their underlying mechanisms are similar. The system can perform some simple profiling of applications and historical migration performance, and it can decide whether to use two or three layers on a case-by-case basis in real time.

\section{Open Issues}

Our current implementation of containers uses the default directory backing store, and as a result, each additional layer duplicates the entire filesystem of the previous layer plus the new data. For a layered setup like the one we demonstrated, a more ideal solution would be to try using an overlay filesystem such as overlayFS, which allows the sharing of lower-layer files with different upper layers. This would provide a much more efficient usage of storage, and is much closer to the abstract model. However, LXC currently does not support overlayFS yet (as of version 1.1.5). 

Another way of potentially improving performance, especially service downtime, would be to investigate the use of iterative migration. Iterative migration is the process of transferring memory pages whilst the service is running, so that when the service is finally suspended and migrated, in theory, only a small portion of the remaining in-memory state needs to be transferred.

For our experiments, LXC offered the greatest ease for running different applications without specialist knowledge in order to set them up. Other container technologies exist, among them Docker (\url{www.docker.com}) which has become the de facto standard for containerization. Future research should investigate whether Docker or other container technologies offer a better overall solution. To answer this question, consideration should be given as to what technology is likely to receive wide-spread adoption from industry, and therefore have the greatest impact. An interesting development to this is the work being carried out by the Open Container Initiative (\url{www.opencontainers.org}) which aims to create an open industry standard around container formats and run-times. Extensive industrial backing and a foundation based on Docker's format and run-times makes this a group to follow.

Besides migrating back-end applications between different MEC servers, another interesting aspect that is worth studying is the ``vertical'' migration of code and data between the mobile device and MEC, to strike a balance among device resource consumption and service quality. This is also known as application/computation offloading \cite{mobileOffload}. The challenge here is that the mobile device's operating system is often different from the server's operating system. A proper encapsulation mechanism that works on both platforms is needed to facilitate the migration. Since the layered framework presented in this article is applicable to a general class of encapsulation methods that supports the suspension of applications, we envision that a similar approach can be applied to live migration  between mobile device and MEC. A detailed study on the vertical migration can be conducted in the future. 

Future work should also study the live migration performance under large-scale networked MEC systems. The performance of migration decision making in large-scale MEC environments has been mainly studied using simulations, where no real application migration is carried out \cite{wang2015dynamic,ceselli2015cloudlet}. We have focused on the other end of the problem in this article, namely the implementation of live migration itself. In the future, it is worthwhile to study the migration of real applications in a realistic, large-scale networked system, which would connect the theoretical results in \cite{wang2015dynamic,ceselli2015cloudlet} with the systems work in this article.

\section{Summary}

We have presented a layered framework for service migration in MECs. The framework supports both container and VM technologies, and it can be easily implemented using existing functionalities of popular container and VM implementations. Extensive experimentation results on the performance of different approaches for various applications under different scenarios have been presented. In general, the three-layer model with a container-based encapsulation environment gives the best overall performance, but other options may be preferred in specific cases, as discussed. The three-layer model also allows the pre-caching of popular applications at MECs, so that the time required for future instantiation of such applications can be shortened. In addition, as migration is performed on the entire container or VM, the underlying service applications do not need to be specifically modified to support migration. This makes it easy to run existing applications in the migration framework. 
A future implementation may also specify an optional interface between the framework and the application, so that the application can announce when it prefers or does not prefer to be migrated, thereby improving the migration performance.

\section*{Acknowledgement}
Some preliminary results only on the three-layer model were presented as a 2-page poster abstract in \cite{machen2016migration}.
This article provides more comprehensive discussion and experimentation results.

Andrew Machen's contribution to this work was performed while he was affiliated with Imperial College London and IBM U.K.

This research was sponsored in part by the US Army Research Laboratory and the UK Ministry of Defense and was accomplished under Agreement Number W911NF-16-3-0001. The views and conclusions contained in this document are those of the author(s) and should not be interpreted as representing the official policies, either expressed or implied, of the US Army Research Laboratory, the US Government, the UK Ministry of Defense or the UK Government. The US and UK Governments are authorized to reproduce and distribute reprints for Government purposes notwithstanding any copyright notation hereon.

\bibliographystyle{IEEEtran}
\bibliography{bibliography}

\end{document}